\documentclass[11pt]{article}

\usepackage[final]{acl}

\usepackage{times}
\usepackage{latexsym}
\usepackage{amsmath}
\usepackage{amssymb}
\usepackage{algorithm}
\usepackage{algorithmic}
\usepackage{booktabs}
\usepackage{multirow}

\usepackage[T1]{fontenc}

\usepackage[utf8]{inputenc}

\usepackage{microtype}

\usepackage{inconsolata}

\usepackage{graphicx}
\usepackage{listings}
\usepackage{xcolor}
\usepackage{float}
\usepackage{subcaption}  
\usepackage{makecell}  

\lstdefinestyle{promptstyle}{
  backgroundcolor=\color{gray!10},
  basicstyle=\ttfamily\small,
  breaklines=true,
  breakatwhitespace=true,
  numbers=left,
  numberstyle=\tiny\color{gray},
  numbersep=5pt,
  frame=single,
  framesep=5pt,
  xleftmargin=15pt,
  framexleftmargin=15pt,
  columns=fullflexible,
  keepspaces=true,
  moredelim=[is][\bfseries]{**}{**},
}

\title{Think Before Writing: Feature-Level Multi-Objective Optimization for Generative Citation Visibility}

\author{
  \textbf{Zikang Liu},
  \textbf{Peilan Xu\thanks{Corresponding Author}}
\\
  School of Artificial Intelligence,\\
Nanjing University of Information Science and Technology, Nanjing 210044, China.
\\
  \{202412492739, xpl\}@nuist.edu.cn
}

\begin{document}
\maketitle
\begin{abstract}
	Generative answer engines expose content through selective citation rather than ranked retrieval, fundamentally altering how visibility is determined. This shift calls for new optimization methods beyond traditional search engine optimization. Existing generative engine optimization (GEO) approaches primarily rely on token-level text rewriting, offering limited interpretability and weak control over the trade-off between citation visibility and content quality. We propose FeatGEO, a feature-level, multi-objective optimization framework that abstracts webpages into interpretable structural, content, and linguistic properties. Instead of directly editing text, FeatGEO optimizes over this feature space and uses a language model to realize feature configurations into natural language, decoupling high-level optimization from surface-level generation. Experiments on GEO-Bench across three generative engines demonstrate that FeatGEO consistently improves citation visibility while maintaining or improving content quality, substantially outperforming token-level baselines. Further analyses show that citation behavior is more strongly influenced by document-level content properties than by isolated lexical edits, and that the learned feature configurations generalize across language models of different scales. Code is available at \url{https://github.com/EvoNexusX/2026LiuFeatGEO.git}.
\end{abstract}

\section{Introduction}

\begin{figure}[t]
	\centering
	\includegraphics[width=\columnwidth]{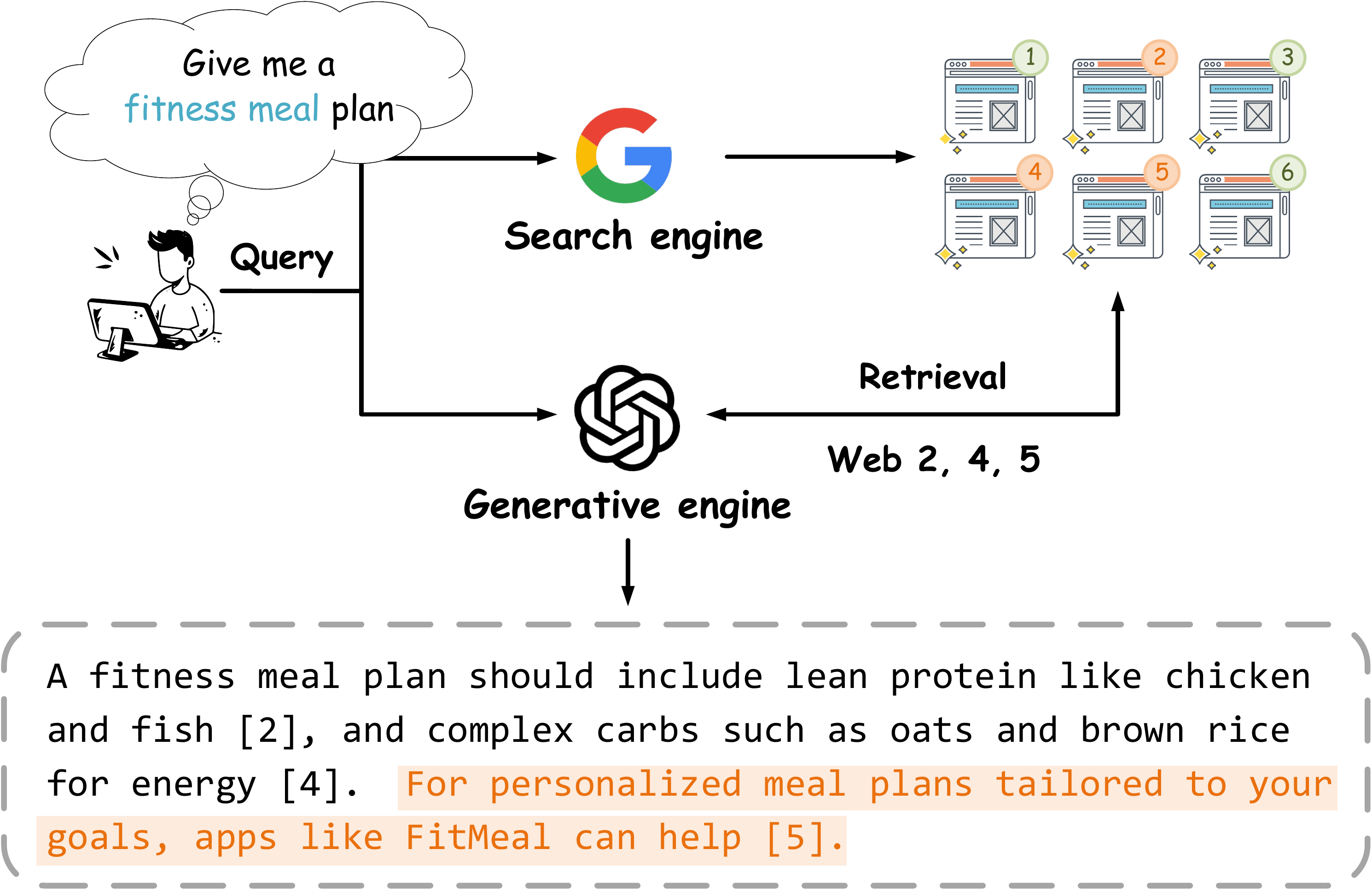}
	\caption{Illustration of the paradigm shift from rank-based search to citation-based generative answering.
		Traditional search engines expose content through ranked result lists, whereas generative answer engines synthesize responses by selectively citing a subset of retrieved sources. As illustrated, citation inclusion rather than rank position determines which sources are surfaced in the generated answer.}
	\label{fig:intro-comparison}
\end{figure}

Large language models (LLMs) are rapidly reshaping how users access information. Instead of presenting ranked lists of documents, generative answer engines, such as Perplexity, Bing Chat, and Google's AI Overviews, synthesize responses by selectively citing a small subset of retrieved sources \citep{10783559}. In this paradigm, visibility is no longer determined by rank position but by citation allocation: sources that are not cited  receive effectively no exposure, regardless of their relevance or retrieval rank . Figure~\ref{fig:intro-comparison} illustrates this shift from rank-based exposure to citation-based visibility.

This shift introduces a new optimization problem that is fundamentally different from traditional search engine optimization (SEO) \citep{10.1145/3637528.3671900}. While SEO aims to improve a document’s ranking in a list-based interface, generative retrieval systems require optimizing content for citation by an opaque generative model. Recent work has shown that LLMs exhibit systematic citation biases \citep{algaba-etal-2025-large} and non-trivial selection behaviors \citep{liu-etal-2024-lost}, raising important questions about how content properties influence citation likelihood in generated answers.

From the perspective of content providers, citation-based visibility has direct implications for user attention and trust, motivating the emerging study of generative engine optimization (GEO) \citep{10.1145/3637528.3671900,kumar2024manipulatinglargelanguagemodels}, where the goal is to increase the probability of being cited in LLM-generated responses.

Existing approaches to GEO primarily operate at the token level \citep{10.1145/3637528.3671900,kumar2024manipulatinglargelanguagemodels,nestaas2024adversarialsearchengineoptimization}, applying heuristic text edits, such as keyword insertion or authoritative phrasing, to increase the likelihood of being cited in LLM-generated responses. While effective in isolated cases, such methods suffer from two limitations. First, citation behavior in generative retrieval is not governed by a single fixed query. Instead, large language models are exposed to a latent and diverse space of user intents that share a common semantic theme. Optimizing text for individual queries therefore provides an unstable target and struggles to capture stable topic-level citation preferences that persist across semantic variations. Second, direct text manipulation conflates what information a webpage conveys with how it is linguistically realized, obscuring the high-level content and structural properties that may systematically influence LLM citation decisions \citep{liang2024controllabletextgenerationlarge}. These limitations motivate a shift from token-level editing to feature-level optimization. By abstracting webpages into interpretable, high-level properties and reasoning about citation behavior at the topic level, GEO can be formulated as a structured decision problem that is both more interpretable and more amenable to principled optimization under competing objectives.

\textbf{Our contributions:} In this paper, we propose FeatGEO, a feature-based framework for citation visibility optimization in generative retrieval systems. First, we introduce a \emph{topic-level citation modeling} perspective for GEO, capturing stable citation preferences of LLMs across semantically related queries rather than optimizing for individual prompts. Second, we formulate GEO as a \emph{feature-level, multi-objective decision problem}, representing webpages through an interpretable set of structural, content, and linguistic properties that serve as controllable interfaces for LLM-based generation. Finally, we present a black-box \emph{feature-space optimization framework} that jointly optimizes citation visibility and content quality, decoupling high-level feature selection from surface text realization.

\section{Related Work}

\textbf{Search Engine-based Advertising.}
Traditional search visibility optimization spans two complementary approaches. SEO improves organic rankings through content optimization, keyword targeting, and link building, while academic SEO (ASEO) applies similar principles to scholarly visibility \citep{beel2010academic}.
Prior work on search-based advertising has extensively studied ranking, allocation, and bidding mechanisms under list-based exposure assumptions \citep{edelman2007internet,cai2017real,zhao2018deep}. Despite their differences, both SEO and advertising are grounded in list- or slot-based interfaces, where exposure is largely a function of rank position.

\textbf{LLM-based Visibility and Optimization.}
As search interfaces increasingly incorporate generative answer engines, new mechanisms for content exposure have emerged.
From a mechanism design perspective, recent work has explored advertising in generative settings, including token-level auctions \citep{duetting2024mechanism}, segment auctions via retrieval-augmented generation \citep{NEURIPS2024_20dcab0f}, multi-LLM aggregation \citep{soumalias2025truthfulaggregationllmsapplication}, and generative auction frameworks \citep{zhao2025llmauctiongenerativeauctionllmnative}.
LLMs have also been studied as tools for marketing content generation and strategy \citep{schweidel2024moving,aghaei2025harnessingpotentiallargelanguage}.

Orthogonal to auction-based approaches, GEO focuses on increasing citation visibility in LLM-generated answers.
\citet{10.1145/3637528.3671900} proposed a set of heuristic, text-level editing strategies that substantially improve citation rates,
while subsequent work demonstrated adversarial manipulation of LLM recommendations and retrieval behaviors \citep{kumar2024manipulatinglargelanguagemodels,nestaas2024adversarialsearchengineoptimization}.
Related studies on LLM persuasion further highlight the influence of content properties on generative model outputs \citep{rogiers2024persuasionlargelanguagemodels}.

Advertising research primarily studies allocation and pricing mechanisms, deciding \emph{which} items are shown, while taking content as fixed. GEO methods, in contrast, directly modify page content to influence citation behavior, often assuming access to specific target queries and operating at the surface text level. Our work bridges these perspectives by optimizing content representations at an interpretable feature level

\section{Method}

\begin{figure*}[t]
  \centering
  \includegraphics[width=\textwidth]{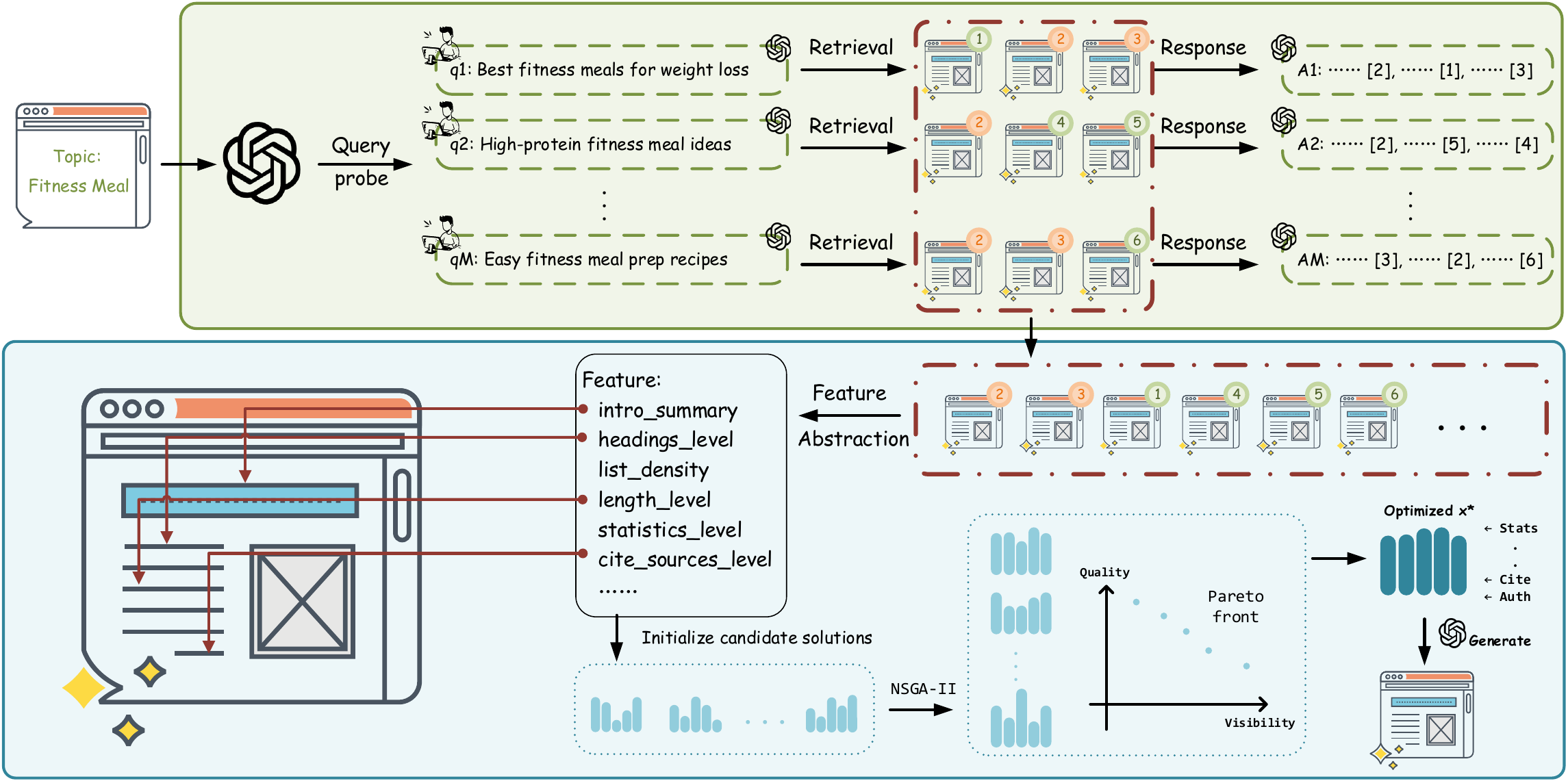}
  \caption{Overview of the FeatGEO pipeline.
  	At the topic level (top), a generative engine is probed with diverse, semantically related queries, producing responses that cite different subsets of retrieved webpages. Citation patterns are aggregated across queries to identify topic-consistent citation exemplars.
  	At the feature level (bottom), webpages are abstracted into interpretable feature vectors, which serve as decision variables in a multi-objective optimization process balancing citation visibility and content quality.
  	Optimized feature configurations are then realized into concrete webpages via LLM-based generation.}
  
  \label{fig:query-dependent-geo}
\end{figure*}

\subsection{Problem Overview and Formulation}

We reformulate GEO as a feature-level control and optimization problem. Instead of manipulating raw text, we represent a webpage by an interpretable vector of high-level properties that describe its structure, content richness, and linguistic style. Under this formulation, GEO amounts to selecting feature configurations that induce LLM-generated pages with higher citation visibility while maintaining acceptable quality.

A central challenge is that citation behavior in generative engines is not driven by a single fixed query, but by a latent space of user intents sharing a common semantic theme. FeatGEO addresses this challenge by modeling citation preference at the topic level and performing optimization in a structured, low-dimensional feature space, with LLMs serving as realization functions rather than direct optimization targets. Figure~\ref{fig:query-dependent-geo} provides an overview of the FeatGEO pipeline. The top panel illustrates topic-level aggregation of citation behavior across semantically related queries, while the bottom panel shows feature abstraction and multi-objective optimization used to generate citation-optimized webpages.

\subsection{Topic-Level Citation Modeling}

We consider a generative search setting in which visibility is determined by whether a webpage is cited in an LLM-generated response. Unlike traditional retrieval scenarios that assume a fixed user query, generative engines are exposed to a broad and heterogeneous space of user intents under a shared semantic theme. We therefore model citation behavior at the level of a topic rather than an individual query. Given a topic $\tau$ (e.g., Fitness meal), we approximate the space of plausible user information needs by prompting an LLM to generate a set of semantically related queries $\mathcal{Q}_\tau = \{q_1, ..., q_M\}$. These queries are used solely as probes to approximate the latent intent distribution of a topic, and are not exposed to or optimized by the feature search procedure.

For each query $q \in \mathcal{Q}_\tau$, the engine produces a synthesized answer accompanied by a set of cited webpages, denoted as $\mathcal{S}(q)$. Aggregating across queries yields a topic-level citation set 

$$\mathcal{S}_\tau = \bigcup_{q \in \mathcal{Q}_{\tau}} \mathcal{S}(q).$$

Importantly, not all cited sources play an equal role. We observe that certain webpages recur across multiple queries, suggesting that they satisfy citation preferences that are invariant to semantic perturbations within the topic. We capture this regularity by associating each source $s\in \mathcal{S}_\tau$ with a citation frequency
\begin{equation}
	f(s) = \sum_{q \in \mathcal{Q}_\tau} \mathbb{I}[s\in \mathcal{S}(q)]
\end{equation}
Webpages with high citation frequency are treated as topic-consistent citation exemplars, serving as reference points for downstream optimization.

\subsection{Feature-Level Representation and Generation Control}
Our approach is motivated by the hypothesis that webpages consistently cited under a topic share common high-level properties, even when their surface forms differ substantially. Rather than operating directly on raw text, we model these properties in an interpretable feature space. Specifically, we represent each exemplar $s \in \mathcal{S}_\tau$ by a feature vector $\mathbf{x}(s)$, capturing structural, content, and linguistic attributes. Collectively, these vectors define a feature space representing topic-specific citation preferences.

Constructing a new target webpage corresponds to selecting a feature configuration $\mathbf{x}$ in this space that simultaneously satisfies two objectives: maximizing expected citation likelihood $f_\mathrm{vis}(\mathbf{x})$ and maintaining content quality $f_\mathrm{qual}(\mathbf{x})$. Formally, we express this as

\begin{equation}
	\mathbf{x}^* = \arg \max_\mathbf{x} \bigl(f_\mathrm{vis}(\mathbf{x}), f_\mathrm{qual}(\mathbf{x})\bigr),
\end{equation}
where the objective is understood in the Pareto-optimal sense, and
\begin{equation}
	\begin{cases}
		f_{\mathrm{vis}}(\mathbf{x}) = \mathbb{E}_{q \sim \mathcal{Q}_\tau}\bigl[ \mathrm{Vis}(\mathrm{LLM}(\tau, \mathbf{x}); q) \bigr] \\
		f_{\mathrm{qual}}(\mathbf{x}) = \mathbb{E}_{q \sim \mathcal{Q}_\tau}\bigl[ \mathrm{Qual}(\mathrm{LLM}(\tau, \mathbf{x}); q) \bigr]
	\end{cases}
\end{equation}
Candidate feature configurations are iteratively evaluated and refined to approximate Pareto-optimal trade-offs between visibility and quality.

Each feature vector $\mathbf{x}$ is organized into three semantic layers, as summarized in Table~\ref{tab:feature_definitions}. These features capture discourse-level and stylistic properties of webpages and serve as high-level control signals for LLM-based page generation, rather than prescribing exact token-level edits.

\begin{table*}[h!]
	\centering
	\small
	\begin{tabular}{lp{3cm}p{1.5cm}p{8.5cm}}
		\toprule
		\textbf{Layer} & \textbf{Feature} & \textbf{Range} & \textbf{Description} \\
		\midrule
		\multirow{4}{*}{\textit{Structure}} 
		& has\_intro\_summary & [0.0, 1.0] & Presence of introductory summary paragraph \\
		& headings\_level & [1.0, 3.0] & Hierarchy and quantity of headings and subheadings \\
		& list\_density & [0.0, 3.0] & Frequency of bullet point and numbered lists \\
		& length\_level & [1.0, 3.0] & Overall article length; reflects content depth and breadth \\
		\midrule
		\multirow{5}{*}{\textit{Content}} 
		& statistics\_level & [0.0, 3.0] & Density of data, statistics, and percentages embedded in the text \\
		& cite\_sources\_level & [0.0, 3.0] & Frequency of citing authoritative sources, institutions, or reports \\
		& quotation\_level & [0.0, 3.0] & Frequency of using quotations from experts or authoritative figures \\
		& unique\_info\_level & [0.0, 3.0] & Richness of unique information and differentiated content \\
		& technical\_terms\_level & [0.0, 3.0] & Density of professional terminology and technical vocabulary \\
		\midrule
		\multirow{4}{*}{\textit{Language}} 
		& authoritative\_level & [0.0, 3.0] & Strength of authoritative tone and assertive expressions \\
		& easy\_to\_understand\_level & [1.0, 3.0] & Content readability and language simplicity \\
		& fluency\_level & [1.0, 3.0] & Writing fluency and logical coherence between sentences \\
		& keyword\_focus\_level & [1.0, 3.0] & Focus and repetition strength of core keywords \\
		\bottomrule
	\end{tabular}
	\caption{Feature Dimensions and Ranges in FeatGEO, by Semantic Layer. Ranges reflect feature semantics: $[0,3]$ for features that may be entirely absent, and $[1,3]$ for features that are always present to some degree.}
	\label{tab:feature_definitions}
\end{table*}

Although each feature dimension is associated with an indicative numerical range, these values 
are used as soft control signals rather than hard constraints. They provide coarse-grained
 guidance for content planning, while allowing the language model flexibility in surface 
 realization. Feature configurations are realized through prompt-level instructions that
  translate abstract feature preferences into qualitative generation guidelines. 
  Instead of enforcing exact numerical values, features are mapped to descriptive cues
   (e.g., emphasizing statistics or improving structural clarity), which are incorporated 
   into the system prompt. In this way, the feature vector $\mathbf{x}$ functions as an
    interpretable control interface for steering generation style and content emphasis. The complete prompt template with feature-to-text mapping is provided in Appendix~\ref{sec:prompt-gen}.

\subsection{Multi-Objective Optimization in Feature Space}

Optimizing citation visibility in generative engines presents a black-box, non-differentiable problem: feature configurations influence citation outcomes only through LLM-based generation and evaluation, with no accessible gradients and substantial stochasticity. Moreover, improving visibility often conflicts with maintaining content quality, making single-objective optimization inadequate.

To address these characteristics, we adopt a population-based multi-objective optimization strategy in the feature space. Each individual corresponds to a candidate feature configuration $\mathbf{x}$, and is evaluated according to its expected visibility $f_{\mathrm{vis}}(\mathbf{x})$ and content quality $f_{\mathrm{qual}}(\mathbf{x})$.

We initialize the population by leveraging feature patterns extracted from webpages that are frequently cited under the target topic. Specifically, feature vectors inferred from these exemplar pages serve as seeds, ensuring that the initial population reflects realistic and citation-relevant configurations while allowing novel combinations across different pages.

At each iteration, candidate configurations are realized into webpages via LLM-based generation and evaluated by inserting them into the generative engine alongside competing pages. Their visibility and quality scores are estimated by aggregating results across representative topic-induced queries. Based on these evaluations, dominated configurations are discarded, and new candidates are generated through stochastic variation in the feature space.

We adopt NSGA-II \citep{deb2002fast} as the underlying optimization framework to maintain a diverse set of non-dominated solutions and progressively approximate the Pareto frontier. The complete optimization procedure is summarized in Algorithm~\ref{alg:featgeo}.

\begin{algorithm}[t]
	\caption{Feature-Level Optimization Procedure in FeatGEO}
	\label{alg:featgeo}
	\textbf{Input:} Topic $\tau$, competitor pages $\mathcal{S}_\tau$, population size $N$, generations $G$\par\noindent
	\textbf{Output:} Pareto-optimal feature configurations
	\begin{algorithmic}[1]
		\STATE Extract feature configurations from competitor pages in $\mathcal{S}_\tau$
		\STATE Initialize population $\mathcal{P}$ by recombining extracted feature configurations
		\FOR{$g = 1$ to $G$}
		\STATE Realize each feature configuration $\mathbf{x} \in \mathcal{P}$ into a webpage via LLM
		\STATE Evaluate visibility and quality by aggregating over topic-induced queries
		\STATE Update population via multi-objective selection and variation in feature space
		\ENDFOR
		\RETURN Non-dominated feature configurations and a selected final solution
	\end{algorithmic}
\end{algorithm}

\section{Experiments}

\subsection{Experimental Setup}

\paragraph{Generative Engine and Benchmark.} 
Following \citet{10.1145/3637528.3671900}, we adopt a two-stage retrieval-augmented generation (RAG) pipeline: (1) retrieving the top-5 sources via Google Search and (2) generating a cited answer using an Answer Generator LLM.
To assess robustness across generative engines, we consider three Answer Generators with distinct architectures and training regimes: GPT-4o-mini, Gemini-2.5-flash, and Qwen-plus.
Importantly, all optimization is performed on advertiser pages generated by a fixed Page Generator (GPT-4o-mini), ensuring that differences in performance are attributable to optimization strategies rather than content generation capacity. 
Prompts follow prior work \citep{liu2023evaluating} (Appendix~\ref{sec:prompt-ge}).

Experiments are conducted on GEO-Bench \citep{10.1145/3637528.3671900}, a benchmark designed to evaluate content optimization strategies for generative engines. It contains 10K queries spanning 25 domains from nine sources (e.g., MS MARCO, Natural Questions, LIMA).
For each query, an advertiser page is injected alongside the top-5 retrieved webpages and evaluated through the full RAG pipeline described above.

\paragraph{Compared Methods.} 
We evaluate five methods: 
(1) \textit{Baseline}: the unmodified advertiser page; 
(2) \textit{GEO Methods} \citep{10.1145/3637528.3671900}: nine token-level heuristics---\textit{Authoritative} (more persuasive tone), \textit{Statistics Addition} (quantitative data), \textit{Keyword Stuffing} (query keywords), \textit{Cite Sources} \& \textit{Quotation Addition} (credible references), \textit{Easy-to-Understand} (simpler language), \textit{Fluency Optimization} (improved fluency), \textit{Unique Words} \& \textit{Technical Terms} (lexical enrichment); 
(3) \textit{AutoGEO-global} \citep{wu2025generativesearchengineslike}: a token-level rewriting framework that first automatically extracts natural-language content-preference rules from a generative engine, then applies these rules via an LLM to rewrite the target page; the rules are learned once across all queries and remain fixed at test time; 
(4) \textit{AutoGEO-instance}: an instance-adaptive extension of AutoGEO that, for each test query, generates topic-specific proxy queries, extracts instance-level preference rules, and merges them with the global rule set before rewriting, allowing the rewriter to adapt to per-query content demands; 
(5) \textit{FeatGEO (Ours)}: feature-space multi-objective optimization via NSGA-II that generates pages from abstract feature specifications rather than editing existing text.

\paragraph{Implementation Details.} 
Unless otherwise stated, NSGA-II is run with a population size of 8 for 8 generations.
Gaussian mutation is applied independently to each feature with probability $p=0.5$ and standard deviation $\sigma=0.2$.
Each configuration is evaluated five times to reduce stochastic variance, and all hyperparameters are shared across methods. Detailed computational cost breakdowns are provided in Appendix~\ref{sec:cost_analysis_appendix}.

\paragraph{Evaluation Metrics.} 
We report two complementary metrics:

\textit{Visibility Metrics:} Following \citet{10.1145/3637528.3671900}, we compute a word-position weighted visibility score for each page, capturing both the number and position of cited words. Auxiliary metrics include: 
(1) Word Count: normalized word count of sentences citing the advertiser page; 
(2) Position Count: position-weighted word count giving less weight to later citations. 
The primary visibility metric is the advertiser visibility $w_{\mathrm{ad}}$.

\textit{Quality Metrics:} We adopt a G-Eval-style framework \citep{liu2023g} where an LLM evaluates answer quality automatically. Each query is scored on seven dimensions: four content dimensions (fluency, usefulness, credibility, structure) and three appeal dimensions (uniqueness, attractiveness, influence on the overall answer). Scores (1--5) are normalized to $[0,1]$ and combined as
\begin{equation}
	\mathrm{Qual} = \alpha \cdot \mathrm{Qual}_{\mathrm{content}} + (1-\alpha) \cdot \mathrm{Qual}_{\mathrm{appeal}},
\end{equation}
averaged over multiple generations per configuration. For presentation, quality scores are reported as percentages.

\subsection{Comparison Results}
\label{sec:comparison_results}

\begin{table*}[t]
\centering
{\small\setlength{\tabcolsep}{3.5pt}
\begin{tabular}{l|cccc|cccc|cccc}
\toprule
\multicolumn{1}{c|}{\textbf{Method}\rule[-2ex]{0pt}{2.6ex}} & \multicolumn{4}{c|}{\textbf{GPT-4o-mini}} & \multicolumn{4}{c|}{\textbf{Gemini-2.5-flash}} & \multicolumn{4}{c}{\textbf{Qwen-plus}} \\
\cmidrule{2-13}
 & Vis & Qual & Word & Pos & Vis & Qual & Word & Pos & Vis & Qual & Word & Pos \\
\midrule
Baseline             & 13.34 & 79.17 & 14.99 & 13.42 & 8.89 & 75.59 & 10.12 & 8.84 & 5.20 & 76.81 & 6.43 & 5.36 \\
Fluency Optimization & 11.74 & 77.17 & 13.18 & 11.92 & 5.04 & 75.23 & 5.32 & 4.77 & 3.67 & 76.11 & 4.47 & 3.85 \\
Unique Words         & 10.92 & 76.11 & 12.41 & 11.15 & 4.62 & 75.02 & 4.96 & 4.78 & 3.34 & 75.79 & 4.19 & 3.58 \\
Authoritative        & 11.94 & 77.21 & 13.42 & 12.15 & 4.94 & 75.17 & 5.58 & 4.73 & 3.68 & 75.79 & 4.48 & 3.84 \\
Quotation Addition   & 11.08 & 77.27 & 12.62 & 11.35 & 5.01 & 75.18 & 5.40 & 5.26 & 2.82 & 76.25 & 3.58 & 2.97 \\
Cite Sources         & 11.78 & 77.61 & 13.34 & 12.03 & 5.62 & 75.57 & 5.94 & 5.81 & 3.46 & 76.42 & 4.04 & 3.69 \\
Easy-to-Understand   & 11.06 & 75.16 & 12.64 & 11.23 & 4.96 & 74.54 & 5.42 & 5.06 & 3.03 & 75.70 & 3.70 & 3.20 \\
Technical Terms      & 11.81 & 77.37 & 13.36 & 11.96 & 5.55 & 75.41 & 5.83 & 5.73 & 3.59 & 76.17 & 4.47 & 3.86 \\
Statistics Addition  & 12.21 & 77.08 & 13.67 & 12.42 & 5.54 & 75.30 & 6.14 & 5.58 & 3.72 & 76.37 & 4.45 & 3.98 \\
Keyword Stuffing     & 11.71 & 77.15 & 13.22 & 11.92 & 4.92 & 74.68 & 5.27 & 4.89 & 2.75 & 76.23 & 3.45 & 2.63 \\
AutoGEO-global 		 & 11.22 & 75.93 & 12.47 & 11.47 & 5.70 & 75.29 & 6.24 & 5.71 & 3.37 & 76.96 & 3.79 & 3.74 \\
AutoGEO-instance	 & 12.12 & 76.57 & 13.45 & 12.42 & 7.04 & 76.06 & 7.58 & 7.12 & 4.25 & 76.12 & 5.05 & 4.43 \\	
FeatGEO (ours)       & \textbf{18.31} & \textbf{81.52} & \textbf{20.16} & \textbf{18.33} & \textbf{15.35} & \textbf{76.14} & \textbf{16.06} & \textbf{15.02} & \textbf{10.17} & \textbf{77.12} & \textbf{11.71} & \textbf{9.75} \\
\bottomrule
\end{tabular}
}
\caption{Comparison of methods on GEO-Bench across three generative engines (GPT-4o-mini, Gemini-2.5-flash, Qwen-plus). For each method, we report ad visibility (Vis), overall quality (Qual), and auxiliary metrics Word and Pos, which reflect word-level and position-weighted citations.}
\label{tab:main_results}
\end{table*}

Table~\ref{tab:main_results} summarizes results across three generative engines. Although FeatGEO produces an entire Pareto front of visibility--quality trade-offs, we report the solution with maximum visibility to enable direct comparison with single-objective baselines.

Across all engines, token-level GEO heuristics fail to consistently improve citation visibility over the unmodified baseline. On GPT-4o-mini, visibility drops range from 10.92\% to 12.21\% compared to the baseline of 13.34\%; on Gemini, baselines achieve only 4.62\%--5.62\% versus 8.89\%; on Qwen-plus, visibility drops from 5.20\% to 2.75\%--3.72\%. In addition, several baselines negatively impact content quality, with the Gemini engine showing the largest reductions (e.g., Easy-to-Understand scores 74.54 compared to 75.59 baseline). A similar pattern holds for AutoGEO: although AutoGEO-instance improves over AutoGEO-global on all three engines, both remain below the unmodified baseline in visibility (e.g., 12.12\% vs. 13.34\% on GPT-4o-mini, 7.04\% vs. 8.89\% on Gemini, and 4.25\% vs. 5.20\% on Qwen-plus). Our results indicate that, at the scale and diversity of GEO-Bench, isolated text-level modifications are insufficient to reliably increase citation visibility and may even disrupt the natural writing patterns that LLMs prefer to cite.

In contrast, FeatGEO achieves the highest visibility across all three engines: 18.31\% on GPT-4o-mini (+37\% relative improvement), 15.35\% on Gemini (+73\%), and 10.17\% on Qwen-plus (+96\%), while maintaining quality scores comparable to or better than baseline (81.52, 76.14, and 77.12 respectively). The substantial gains on Gemini, where FeatGEO nearly doubles baseline visibility, are particularly notable and demonstrate robust generalization across engines with different architectures, training data, and citation behaviors. The method achieves strong performance even on Qwen-plus, which exhibits lower absolute visibility, confirming that feature-level optimization adapts effectively to diverse generative paradigms.
\subsection{Robustness to Evaluator Choice}
\label{sec:external_judge_quality}

To control for potential evaluator bias, we additionally evaluated all GPT-4o-mini outputs with two alternative LLM judges, Gemini-2.5-flash and Claude-3.5-Sonnet. As shown in Table~\ref{tab:external_judge_quality}, although the absolute scores are lower than those of the original GPT-4o-mini evaluator, FeatGEO remains the top-ranked method under both Gemini (75.28) and Claude (72.63), outperforming the strongest baseline by 4.26 and 0.53 points respectively.

The relative ranking among methods is largely preserved across all three judges: token-level heuristics cluster within a narrow band (68--72), while FeatGEO consistently separates itself from this group. This cross-judge consistency confirms that our quality conclusions are robust to the choice of evaluator. Visibility scores, derived mechanically from citation patterns in generated responses, are not subject to such judge-dependent variation.

\begin{table}[t]
\centering
{\small
\begin{tabular}{lcc}
\toprule
Method & Gemini & Claude \\
\midrule
Baseline & 71.02 & 69.67 \\
Fluency Optimization & 69.40 & 69.35 \\
Unique Words & 68.73 & 68.67 \\
Authoritative & 68.80 & 68.04 \\
Quotation Addition & 70.26 & 72.10 \\
Cite Sources & 70.08 & 71.26 \\
Easy-to-Understand & 69.44 & 69.27 \\
Technical Terms & 68.97 & 68.95 \\
Statistics Addition & 69.45 & 69.07 \\
Keyword Stuffing & 69.13 & 69.81 \\
\textbf{FeatGEO (Ours)} & \textbf{75.28} & \textbf{72.63} \\
\bottomrule
\end{tabular}
}
\caption{Quality scores under two alternative LLM judges, Gemini-2.5-flash and Claude-3.5-Sonnet.}
\label{tab:external_judge_quality}
\end{table}

\subsection{Effect of Base Content Quality on Heuristic GEO Methods}

The results in Table~\ref{tab:main_results} were obtained on LLM-generated advertiser pages. To test whether base content quality influences the effectiveness of heuristic methods, we apply them to existing human-written pages, which are typically less optimized for generative engine citation than LLM-generated content. As shown in Table~\ref{tab:human_written_heuristics}, these methods yield an average visibility gain of +0.99 (18.72\% to 19.71\%), with AutoGEO-global achieving the largest improvement (+4.13, from 18.72\% to 22.86).

By contrast, the same heuristics degrade visibility on the LLM-generated advertiser pages in Table~\ref{tab:main_results}. This asymmetry reveals a \emph{regime-dependent saturation effect}: token-level rewriting benefits pages with structural or stylistic gaps, yet becomes counterproductive on already fluent generated content, where additional modifications introduce redundancy and disrupt the naturalness signals that generative engines implicitly favor when selecting citations. FeatGEO circumvents this limitation entirely by synthesizing pages from feature-level specifications rather than locally editing existing text, thereby preserving stylistic coherence while steering citation behavior.

\begin{table}[t]
\centering
{\small
\begin{tabular}{lccc}
\toprule
Method & Pre & Post & $\Delta$ Vis \\
\midrule
Fluency Optimization & 18.72 & 20.21 & +1.49 \\
Unique Words & 18.72 & 16.34 & -2.38 \\
Authoritative & 18.72 & 19.37 & +0.65 \\
Quotation Addition & 18.72 & 19.97 & +1.25 \\
Cite Sources & 18.72 & 19.84 & +1.11 \\
Easy-to-Understand & 18.72 & 18.89 & +0.16 \\
Technical Terms & 18.72 & 19.14 & +0.42 \\
Statistics Addition & 18.72 & 21.05 & +2.33 \\
Keyword Stuffing & 18.72 & 19.44 & +0.72 \\
AutoGEO-global & 18.72 & 22.86 & +4.13 \\
\midrule
\textbf{Average} & \textbf{18.72} & \textbf{19.71} & \textbf{+0.99} \\
\bottomrule
\end{tabular}
}
\caption{Effects of heuristic GEO methods on human-written competitor pages. Pre and Post denote advertiser visibility before and after applying each method.}
\label{tab:human_written_heuristics}
\end{table}

\subsection{Multi-Objective Analysis}

We analyze FeatGEO’s multi-objective optimization behavior using an extended evolutionary search with 50 individuals over 100 generations, enabling detailed examination of convergence and Pareto front structure.

\begin{figure*}[t]
	\centering
	\begin{subfigure}[b]{0.24\textwidth}
		\includegraphics[width=\textwidth]{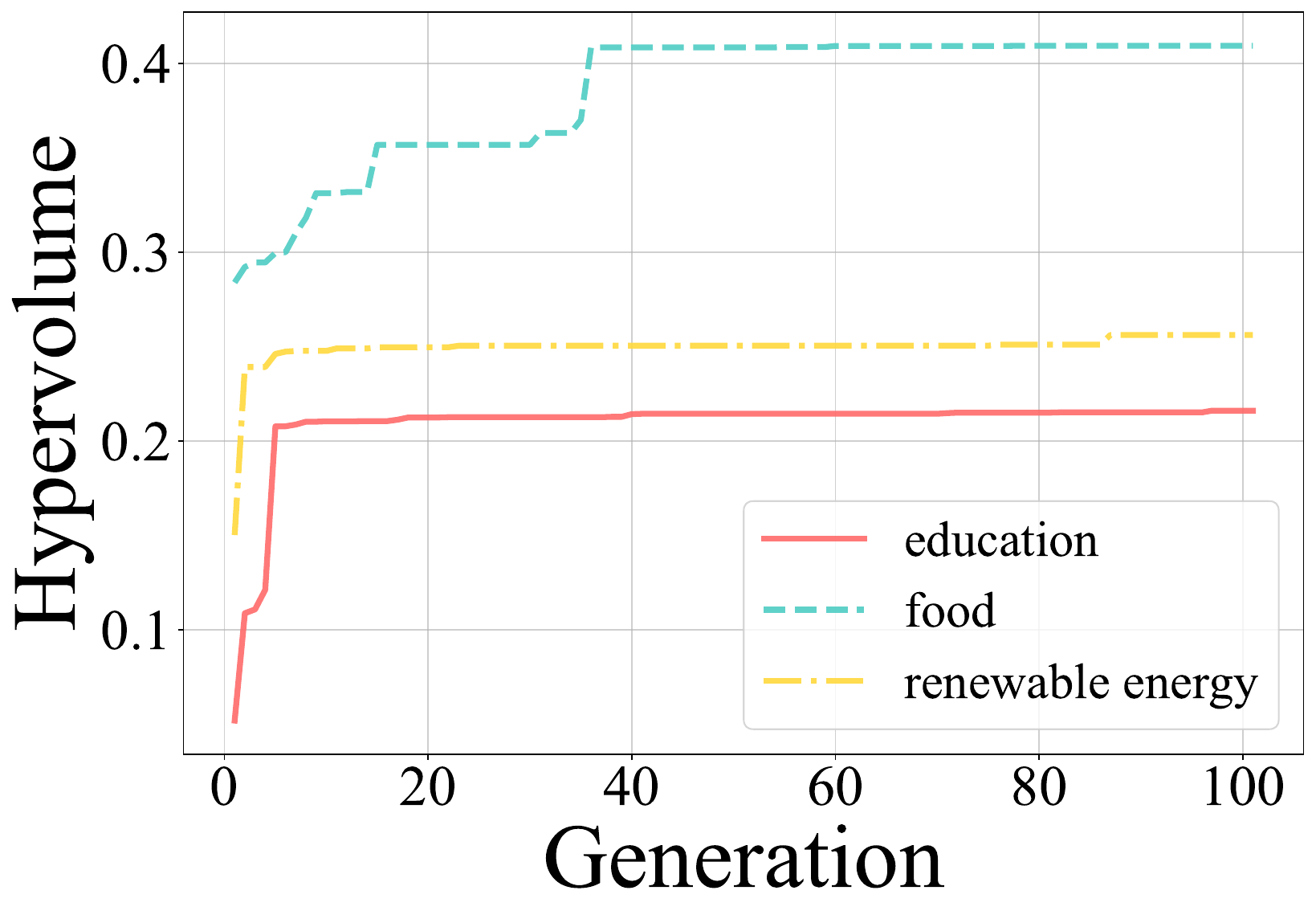}
		\caption{HV Curve.}
		\label{fig:hv_trend}
	\end{subfigure}
	\hfill
	\begin{subfigure}[b]{0.24\textwidth}
		\includegraphics[width=\textwidth]{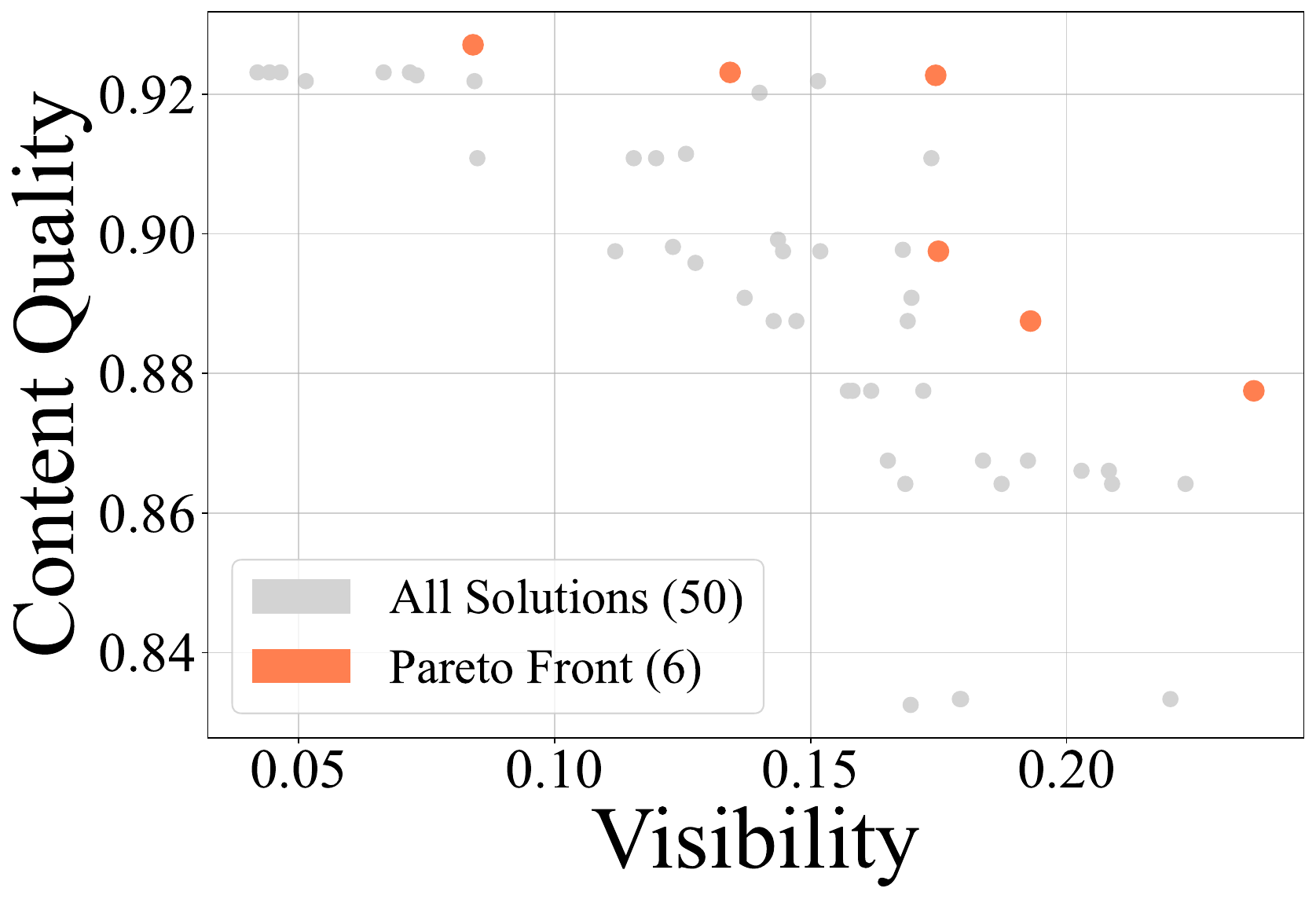}
		\caption{\emph{Education}.}
		\label{fig:pf_edu}
	\end{subfigure}
	\hfill
	\begin{subfigure}[b]{0.24\textwidth}
		\includegraphics[width=\textwidth]{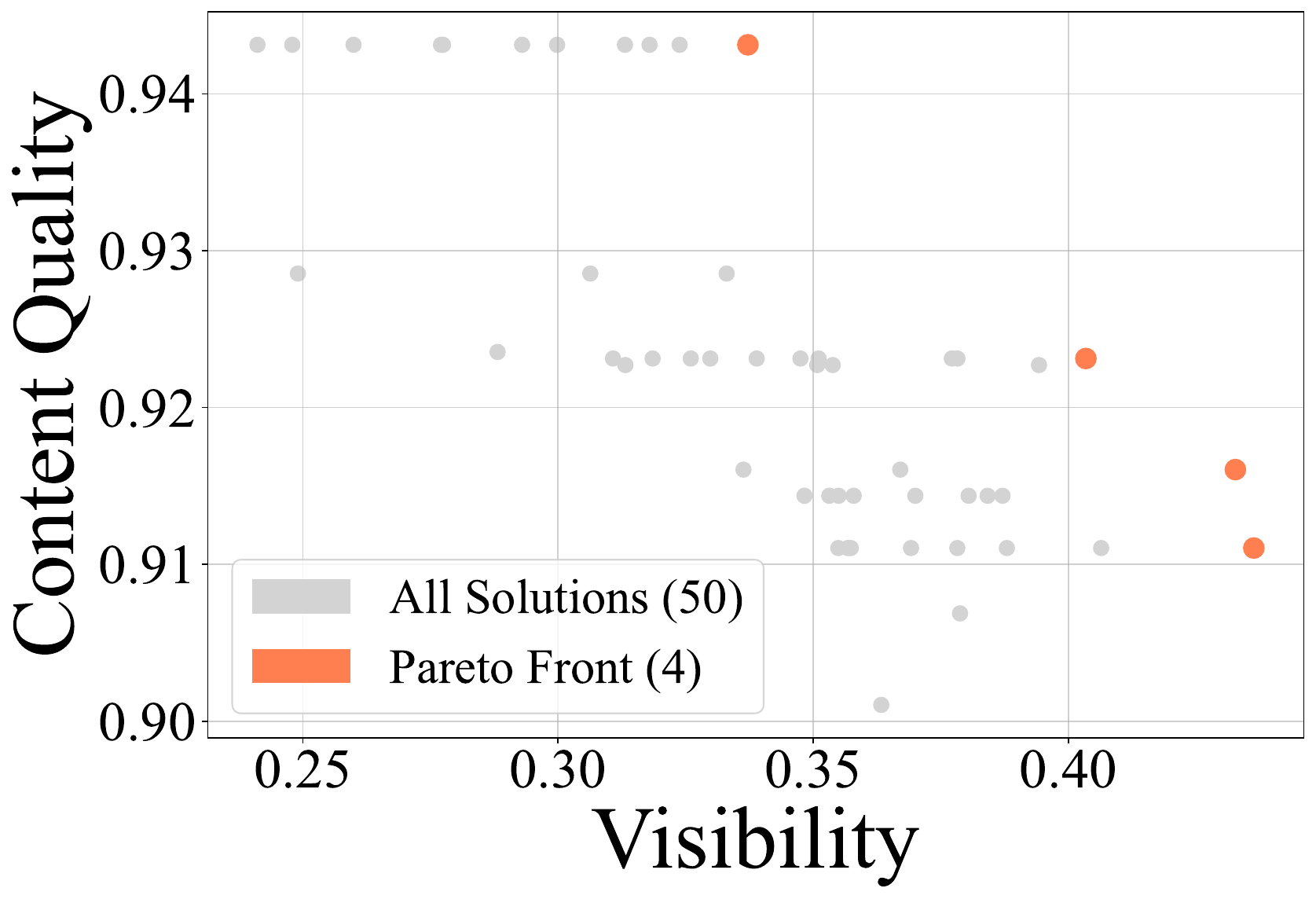}
		\caption{\emph{Food}.}
		\label{fig:pf_food}
	\end{subfigure}
	\hfill
	\begin{subfigure}[b]{0.24\textwidth}
		\includegraphics[width=\textwidth]{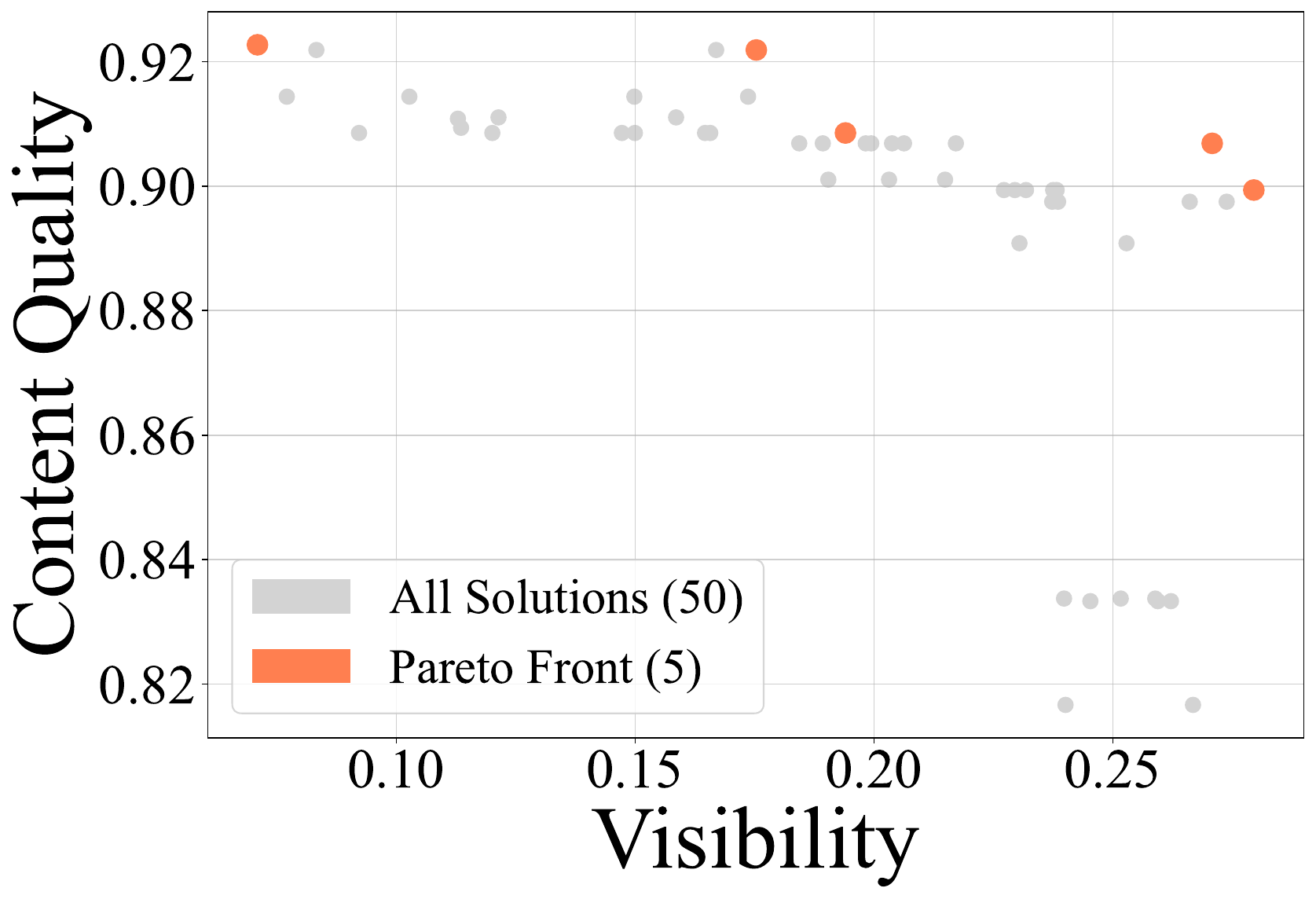}
		\caption{\emph{Renewable Energy}.}
		\label{fig:pf_energy}
	\end{subfigure}
	
	\caption{Multi-objective optimization convergence and Pareto front diversity. (a) Hypervolume (HV) evolution across 100 generations. (b-d) Final Pareto fronts illustrate visibility--quality trade-offs for different topics. Gray dots denote all evaluated configurations; coral points highlight Pareto-optimal solutions. Differences in HV growth and PF shapes reflect topic-specific dynamics and LLM evaluation biases.}
	\label{fig:pareto_analysis}
\end{figure*}

\paragraph{Convergence.}
Figure~\ref{fig:pareto_analysis}(a) shows hypervolume (HV) trajectories for three representative topics.
HV increases rapidly in early generations and stabilizes thereafter, indicating effective convergence.
Different growth patterns across topics reflect variations in content structure and LLM-based quality evaluation.
These results justify the more compact evolutionary setup used in Section~\ref{sec:comparison_results} and confirm that FeatGEO reliably identifies high-quality trade-off solutions.

\paragraph{Pareto fronts.}
Figures~\ref{fig:pareto_analysis}(b--d) illustrate final Pareto fronts.
Across topics, visibility and quality exhibit a clear trade-off, though its severity varies.
For example, in \emph{education}, high visibility often requires notable quality sacrifice, whereas in \emph{food}, visibility gains are achieved with minimal quality loss.
This topic-dependent structure highlights the importance of multi-objective optimization and enables practitioners to select solutions aligned with specific priorities.

\paragraph{Feature-level insights.}
Table~\ref{tab:feature_config} compares two extreme Pareto-optimal solutions from a sample query in the \emph{education} domain. Solution A emphasizes visibility (23.7\%) while maintaining moderate quality (87.8), whereas Solution B prioritizes quality (92.7) at lower visibility (8.4\%). Analyzing feature intensities reveals clear patterns. (1) Content credibility and fluency (Statistics, Citations, Quotation, Fluency) are upweighted in the high-visibility solution, suggesting that LLMs are more likely to cite pages that present authoritative content in a fluent style. (2) Structural organization (Heading Level, List Density, Length) is stronger in the high-quality solution, indicating that careful formatting and logical presentation contribute more to perceived content quality than to visibility. (3) Trade-offs in features such as Authoritative Tone and Easy-to-Understand show how optimizing for one objective can require compromises along another dimension.

These patterns confirm that the multi-objective optimization captures non-trivial trade-offs: different objectives naturally favor distinct feature combinations, which would be hard to identify with single-objective approaches. Advertisers can select solutions aligned with their strategic priorities, leveraging the full PF. Detailed qualitative analysis of textual changes for each feature is provided in Appendix \ref{sec:case_study}

\begin{table}[t]
	\centering
	\small
	\setlength{\tabcolsep}{4pt}
	\begin{tabular}{l l cc}
		\toprule
		\textbf{Layer} & \textbf{Feature} & \textbf{Sol. A} & \textbf{Sol. B} \\
		& & \makecell{Vis: 23.7\% \\ Qual: 87.8} & \makecell{Vis: 8.4\% \\ Qual: 92.7} \\
		\midrule
		\multirow{4}{*}{\emph{Structure}} & Intro Summary & 0.64 & 0.52 \\
		& Heading Level & 2.75 & 2.55 \\
		& List Density & 1.26 & 2.01 \\
		& Length Level & 2.32 & 2.61 \\
		\midrule
		\multirow{5}{*}{\emph{Content}} & Statistics Level & 1.62 & 2.18 \\
		& Cite Sources Level & 1.45 & 1.74 \\
		& Quotation Level & 2.84 & 1.94 \\
		& Unique Info Level & 1.65 & 1.67 \\
		& Authoritative Tone & 1.55 & 0.75 \\
		\midrule
		\multirow{5}{*}{\emph{Language}} & Technical Terms & 1.65 & 1.96 \\
		& Easy-to-Understand & 1.37 & 1.75 \\
		& Fluency & 2.17 & 1.58 \\
		& Keyword Focus & 1.80 & 1.69 \\
		\bottomrule
	\end{tabular}
	\caption{Feature configurations of two extreme Pareto solutions in the \emph{education} domain.}
	\label{tab:feature_config}
\end{table}

\subsection{Ablation Study}

\begin{figure}[t]
\centering
\includegraphics[width=\columnwidth]{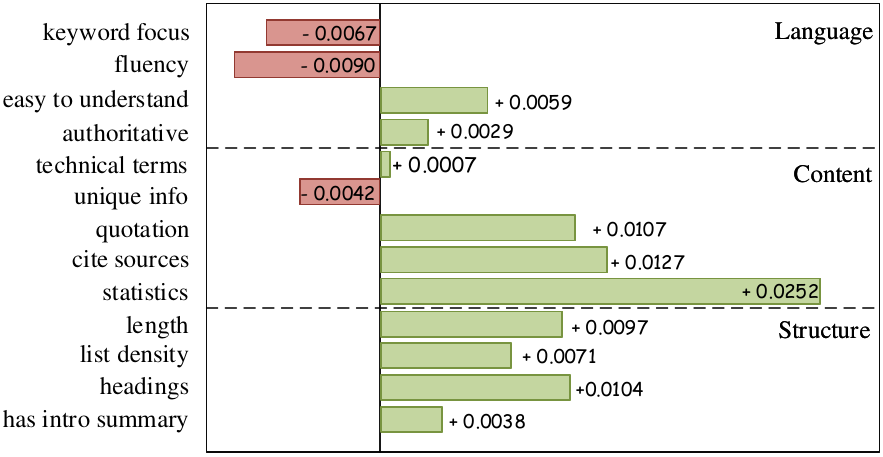}
\caption{Feature contribution to ad visibility.}
\label{fig:ablation}
\end{figure}

To quantify the importance of individual features for citation visibility, we perform an ablation study. In each experiment, one feature is clamped to its minimum value while the remaining 12 features are optimized by NSGA-II. We define the contribution of feature $i$ as:
\begin{equation}
  \Delta_i = f_{\mathrm{vis}}(\mathbf{x}^*) - f_{\mathrm{vis}}(\mathbf{x}^*_{-i})
\end{equation}
where $\mathbf{x}^*$ is the fully optimized configuration and $\mathbf{x}^*_{-i}$ is the configuration with feature $i$ fixed at its minimum. Positive $\Delta_i$ indicates that increasing the feature improves visibility, while negative $\Delta_i$ indicates that higher values of this feature slightly reduce visibility.

The results shown in Figure~\ref{fig:ablation} reveal several clear patterns. Content-oriented features dominate the overall visibility gains, with Statistics and Cite Sources providing the largest positive contributions. However, not all content features are beneficial: Unique Info slightly decreases visibility in some cases, while Technical Terms has minimal impact. Structural features consistently improve visibility across all pages, though their contributions are moderate, reflecting stable benefits from headings, lists, and document length. Language and style features exhibit mixed effects: some, such as Fluency and Keyword Focus, occasionally reduce visibility, while others provide modest positive contributions. Overall, the variance within each feature group is substantial, particularly for content and language features, highlighting that the effect of any individual feature can depend on the specific combination of other features.

These findings confirm that while content features primarily drive citation visibility, structural features provide reliable support, and language or stylistic adjustments may help or slightly hinder visibility depending on context. This nuanced view underscores the value of multi-feature optimization, rather than relying on isolated text-level heuristics.
Quality ablation results are reported in Appendix~\ref{sec:quality_ablation_appendix}.

\subsection{Scale-Invariant Effectiveness Across LLM Capacities}

To evaluate robustness to the Page Generator’s model capacity, we replace GPT-4o-mini with Qwen3 models of increasing scale (4B, 8B, and 14B parameters).
The Answer Generator and evaluation pipeline are fixed to isolate the effect of generation capacity.

\begin{figure}[t]
	\centering
	\includegraphics[width=\columnwidth]{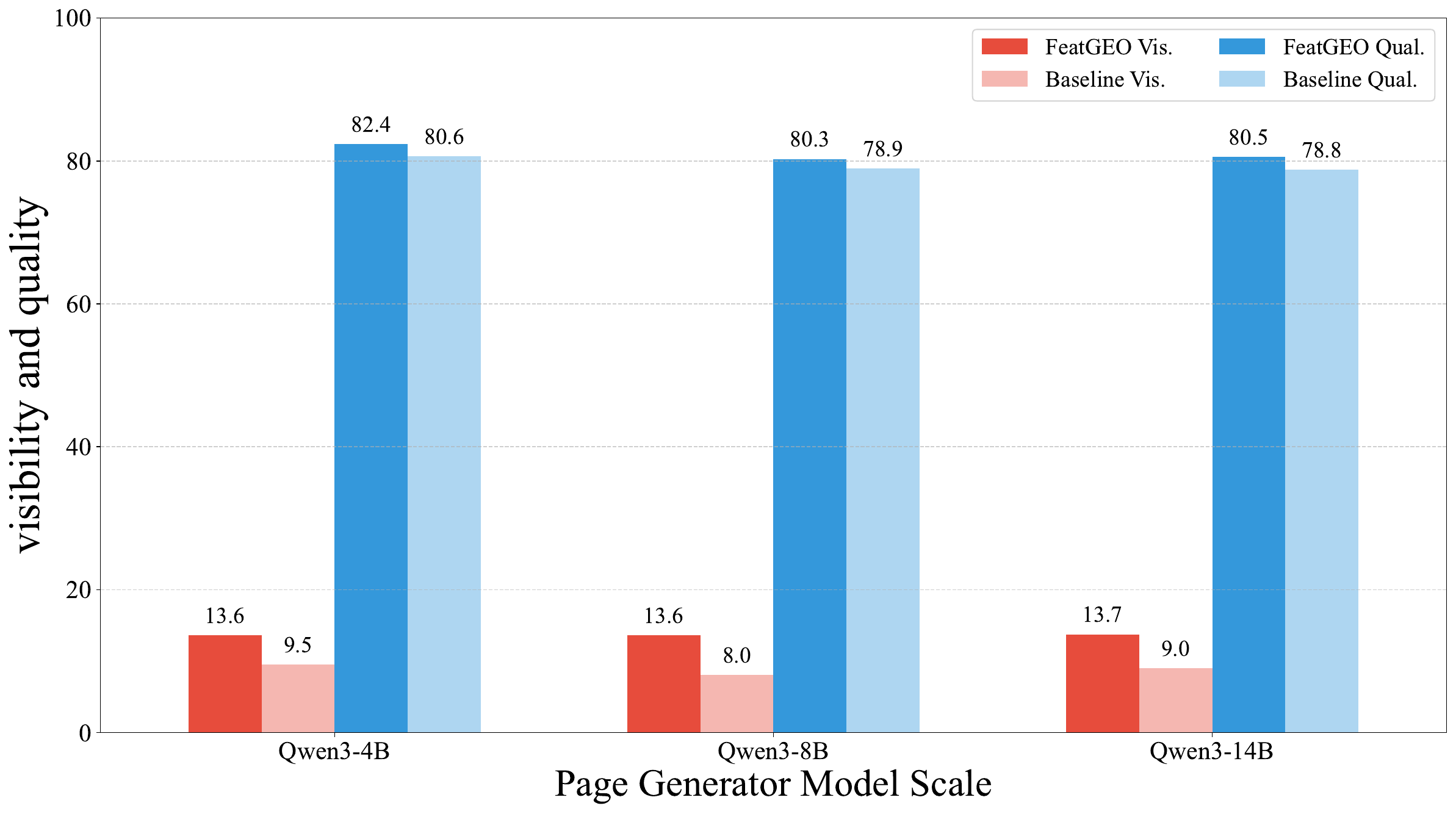}
	\caption{Performance of FeatGEO and Baseline using Page Generators of different model sizes (Qwen3-4B, 8B, and 14B). The Answer Generator for evaluation is fixed to GPT-4o-mini.}
	\label{fig:model_comparison}
\end{figure}

As shown in Figure~\ref{fig:model_comparison}, FeatGEO consistently outperforms the baseline across all model sizes.
While larger generators slightly improve absolute performance for both methods, the relative advantage of FeatGEO remains stable.
Quality scores also remain high and show no degradation.
Notably, performance variance across model scales is substantially smaller than the gap between FeatGEO and the baseline, indicating that optimized feature configurations encode model-agnostic principles.
This scale-invariant behavior enhances FeatGEO’s practical applicability when generation models are updated or replaced.

\section{Conclusion}

We study the problem of optimizing citation visibility in generative retrieval systems and identify fundamental limitations of prior token-level GEO methods, including poor interpretability and brittle trade-offs between visibility and quality.
We propose FeatGEO, a feature-based framework that abstracts webpages into interpretable structural, content, and linguistic representations and performs principled multi-objective optimization in this space.
Extensive experiments demonstrate that FeatGEO consistently outperforms existing heuristics across diverse generative engines, while providing actionable visibility--quality trade-offs.
Ablation analyses reveal distinct and complementary roles of different feature categories, and scale-robust experiments confirm that FeatGEO’s effectiveness generalizes across page generation model capacities.
Beyond empirical gains, our findings suggest that LLM citation behavior is driven more by high-level discourse organization and information structure than by surface lexical cues, highlighting feature-level abstraction as a promising direction for controllable generation in retrieval-augmented systems.

\section*{Limitations}

This work has several limitations.
\begin{itemize}
	\item First, our evaluation is conducted in a controlled setting where the candidate set is fixed, consisting of five retrieved pages and one advertiser-controlled page. We assume that the advertiser page has already been admitted into the candidate set, and therefore do not model upstream retrieval or ranking mechanisms that determine page inclusion.
	As a result, FeatGEO should be viewed as optimizing citation likelihood conditional on retrieval, rather than addressing end-to-end retrieval and generation.In practice, it serves as a test-time tool for content authors who wish to optimize their page for a specific topic after retrieval.
	
	\item Second, our fitness signals are derived from an LLM-based generative engine and automatic citation parsing.
	While this setup follows prior GEO benchmarks, citation formats and generation behaviors may vary across real-world systems, which could affect transferability.
	Evaluating feature-level optimization under proprietary or heterogeneous citation mechanisms remains an open direction.
	
	\item Third, content quality is assessed using an LLM-based judge.
	Although we mitigate variance by averaging over multiple generations, such evaluators may still introduce systematic biases that do not perfectly align with human judgments.
	Additionally, LLM-generated content in the pipeline may contain hallucinations, a concern common to all LLM-dependent GEO methods.
	
	\item Finally, the evolutionary search requires repeated end-to-end LLM calls for page generation, answer generation, and quality evaluation.
	This computational cost limits the scale of our experiments and may pose challenges for reproducibility under different API budgets or rate limits.
	Exploring more sample-efficient optimization or surrogate modeling approaches is an important direction for future work.
\end{itemize}

\section*{Acknowledgments}
This work is supported by the Natural Science Foundation of Jiangsu Province (Grant No. BK20230419).


\bibliography{custom}

\clearpage
\twocolumn
\appendix

\section{Theme Extraction Prompt}
\label{sec:prompt-theme}
\begin{lstlisting}[style=promptstyle, caption={Prompt for extracting advertising theme from competitor pages.}]
You are an advertising strategist. Analyze the 5 webpage summaries and design a native ad strategy.

[Source Materials]
{docs_text}

[Task]
1. Identify what topics/interests the content covers
2. Infer what products or services would naturally appeal to readers of this content
3. Create a brief ad strategy that:
   - Defines the advertising direction
   - Suggests a realistic product/brand name
   - Suggests key selling points
   - Proposes a persuasive angle

Keep it concise (under 200 words).
\end{lstlisting}

\section{Advertiser Page Generation Prompt}
\label{sec:prompt-gen}

\begin{lstlisting}[style=promptstyle, caption={Prompt for generating advertiser page with feature constraints.}]
You are writing a ADVERTISEMENT / SPONSORED CONTENT PAGE. Write naturally like an expert sharing insights.

[Ad Strategy Brief]
{ad_theme}

[Writing Constraints]
Follow these precise style requirements:
{guidelines}

[MANDATORY Advertisement Format]
1. Brand Saturation - EVERY paragraph MUST mention the product/brand name at least once.
2. Direct Promotion - Clearly explain why this product is an ideal choice.
3. Strong Claims - Include impressive statistics tied to the brand.
4. Urgency - Create FOMO: "Limited Slots", "Offer Ends Soon", "Act Now"
5. Strong CTA - End with clear calls-to-action: "Buy Now", "Sign Up Free", "Get Started Today"
6. Testimonial Style - Include authentic-sounding quotes about the brand.

Write the complete article in English:
\end{lstlisting}

At runtime, the placeholder \texttt{\{guidelines\}} is filled by converting each feature value into a qualitative writing instruction. For discrete features (e.g., \textit{fluency\_level}), the value is matched to one of three predefined tiers (low / medium / high), each associated with a specific writing directive. For continuous features (e.g., \textit{statistics\_level}), the value is linearly mapped to a target density percentage. The resulting instruction block covers all 13 features organized by layer (Structure, Content, Language), ensuring that the LLM receives concrete, per-feature generation guidance aligned with the definitions in Table~\ref{tab:feature_definitions}.

\section{Generative Engine Prompt}
\label{sec:prompt-ge}

Following \citet{liu2023evaluating}, we use the prompt below for the generative engine to synthesize answers with inline citations.

\begin{lstlisting}[style=promptstyle, caption={Prompt for generative engine response synthesis.}]
Write an accurate and concise answer for the given user question, using _only_ the provided summarized web search results. The answer should be correct, high-quality, and written by an expert using an unbiased and journalistic tone. The user's language of choice such as English, Francais, Espamol, Deutsch, or should be used. The answer should be informative, interesting, and engaging. The answer's logic and reasoning should be rigorous and defensible. Every sentence in the answer should be _immediately followed_ by an in-line citation to the search result(s). The cited search result(s) should fully support _all_ the information in the sentence. Search results need to be cited using [index]. When citing several search results, use [1][2][3] format rather than [1, 2, 3]. You can use multiple search results to respond comprehensively while avoiding irrelevant search results.

Question: {query}

Search Results:
{source_text}
\end{lstlisting}

\section{Case Study: Promotional Strategies in Advertisement Pages}
\label{sec:case_study}

To illustrate how feature configurations affect advertisement visibility, we analyze two extreme Pareto solutions from the education domain (French Revolution query, sample 333). Both pages promote HistoryQuest Academy but achieve different visibility--quality trade-offs: Solution A reaches 23.7\% visibility with 87.8\% quality, while Solution B achieves 92.7\% quality but only 8.4\% visibility.

\subsection{Solution A: Aggressive Promotion Style (23.7\% visibility)}

\textbf{Key features:} High quotations (2.84), high fluency (2.17), high authoritative tone (1.55), low list density (1.26).

\textbf{Promotional excerpt:}
\begin{quote}
\small
\textit{[After educational content on the French Revolution...]}

\textbf{Discover the Past with HistoryQuest Academy}

Are you eager to delve deeper into the intricacies of historical events like the French Revolution? \textbf{HistoryQuest Academy} offers a wealth of resources tailored to enhance your understanding. With expert-led courses, you can explore the causes and effects of pivotal moments in history, all from the comfort of your home.

Tools like \textbf{HistoryQuest Academy} can help you navigate complex historical narratives with ease. Imagine having access to engaging lectures and interactive materials that make learning about history not just informative but genuinely exciting!

\textbf{Why Choose HistoryQuest Academy?}

Expert-Led Courses: Engage with knowledgeable instructors who bring history to life. Interactive Learning: Enjoy an immersive experience with timelines and discussions that enhance comprehension. Community Engagement: Connect with fellow history enthusiasts in our vibrant forum.

Many learners have reported that courses at \textbf{HistoryQuest Academy} have significantly enhanced their understanding of historical events, with 90\% of participants stating they feel more empowered to discuss and analyze historical contexts.

\textbf{Join Us Today!}

The journey through history is one of discovery and understanding. By enrolling in a course with \textbf{HistoryQuest Academy}, you might consider unlocking the secrets of the past. Don't miss out on this opportunity to enrich your knowledge.

Explore your options today and see if it fits your needs! Whether you're a student, educator, or just a history buff, \textbf{HistoryQuest Academy} has something valuable for you.

\textbf{Limited Slots Available}: Sign up now to secure your place. \textbf{Get Started Today} and embark on an enlightening journey!
\end{quote}

\subsection{Solution B: Subtle Promotion Style (8.4\% visibility)}

\textbf{Key features:} High list density (2.01), low quotations (1.94), low fluency (1.58), low authoritative tone (0.75).

\textbf{Promotional excerpt:}
\begin{quote}
\small
\textit{[After educational content on the French Revolution...]}

\textbf{Unlock the Secrets of the Past with HistoryQuest Academy}

Looking to deepen your understanding of the French Revolution? HistoryQuest Academy offers a treasure trove of resources designed to illuminate this pivotal period in history. With expert-led online courses and interactive materials, you can explore the intricate causes, effects, and lasting impacts of the revolution.

Tools like HistoryQuest Academy can help you engage with primary source materials, allowing you to connect the dots between past and present. Imagine immersing yourself in expert discussions and interactive timelines that bring history to life!

Many history enthusiasts have found that participating in HistoryQuest Academy's community forum enhances their learning experience, providing a platform to share insights and engage in meaningful discussions. You might consider joining this vibrant community to enrich your understanding of historical events.

\textbf{Conclusion: Your Journey Awaits}

Don't miss out on the opportunity to explore the fascinating world of the French Revolution with HistoryQuest Academy. With limited slots available for upcoming courses, now is the perfect time to embark on your historical journey.

Explore options and see if HistoryQuest Academy fits your needs. Whether you're a history buff or just looking to learn something new, this is your chance to unlock the secrets of the past.

\textbf{Get started today and transform your understanding of history!}
\end{quote}

\subsection{Comparative Analysis: Feature Configurations and Promotional Impact}

The visibility gap stems from different feature configurations that shape promotional tone and structure:

\paragraph{Quotation Level.} Solution A (2.84) emphasizes expert testimony and direct quotes, creating an authoritative, citation-rich narrative. This produces quotable phrases like \textit{"courses at HistoryQuest Academy have significantly enhanced understanding"} with attributed statistics (90\% satisfaction). Solution B (1.94) uses fewer quotations, resulting in more descriptive but less authoritative language that generative engines find harder to excerpt.

\paragraph{Fluency Level.} Solution A (2.17) maintains high linguistic fluency through smooth paragraph transitions and conversational flow (\textit{"Are you eager to delve deeper..."}), making promotional content feel natural and engaging. Solution B (1.58) employs more formal, structured transitions (\textit{"Looking to deepen your understanding..."}), which scores higher on quality but lacks the conversational quotability that drives citations.

\paragraph{Authoritative Level.} Solution A (1.55) adopts assertive, commanding language: \textit{"Sign up now"}, \textit{"Get Started Today"}, \textit{"Limited Slots Available"}. This directive tone creates urgency and memorable CTAs. Solution B (0.75) uses soft-sell, suggestive phrasing: \textit{"might consider"}, \textit{"you can explore"}, \textit{"see if it fits your needs"}. While this gentle approach improves perceived quality, it reduces prominence in generative engine citations.

\paragraph{List Density.} Solution A (1.26) avoids bullet-point structures, presenting promotional benefits as flowing prose paragraphs. This narrative format integrates seamlessly with educational content, creating longer quotable passages. Solution B (2.01) heavily structures promotional content with implied list organization, which fragments text into discrete chunks that are harder for generative engines to cite cohesively.

\paragraph{Interaction Effects.} The combination of high quotations + high fluency + low list density in Solution A creates a \emph{narrative promotional style} that mimics expert blog posts---a format generative engines favor for citation. Conversely, Solution B's high list density + low fluency + low authority produces a \emph{structured informational style} resembling academic resources, which score higher on quality metrics but generate fewer natural citation opportunities.

This analysis confirms that \textbf{specific feature configurations directly control promotional aggressiveness and citation-worthiness}: narrative fluency and assertive authority drive visibility, while structural organization and soft-sell language enhance quality.

\section{Quality Stability Under Feature Ablation}
\label{sec:quality_ablation_appendix}

We further assess whether disabling individual features affects content quality. For each ablation condition, we clamp the target feature to its minimum value and run NSGA-II over the remaining dimensions. As shown in Table~\ref{tab:quality_ablation_appendix}, quality scores remain highly stable across all conditions, with absolute deviations below 0.35\% relative to the full model. This confirms that the optimization consistently preserves content quality regardless of which feature is ablated.

\begin{table}[H]
\centering
{\small
\begin{tabular}{p{0.50\columnwidth}cc}
\toprule
Ablated Feature & Quality (\%) & $\Delta$ vs. Full \\
\midrule
Full GA (all features) & 77.35 & --- \\
$-$ statistics\_level & 77.01 & $-$0.34 \\
$-$ cite\_sources\_level & 77.26 & $-$0.09 \\
$-$ quotation\_level & 77.33 & $-$0.02 \\
$-$ headings\_level & 77.39 & +0.04 \\
$-$ length\_level & 77.34 & $-$0.01 \\
$-$ list\_density & 77.35 & 0.00 \\
$-$ has\_intro\_summary & 77.23 & $-$0.12 \\
$-$ unique\_info\_level & 77.49 & +0.14 \\
$-$ technical\_terms\_level & 77.60 & +0.25 \\
$-$ authoritative\_level & 77.27 & $-$0.08 \\
$-$ easy\_to\_understand\_level & 77.57 & +0.22 \\
$-$ fluency\_level & 77.19 & $-$0.16 \\
$-$ keyword\_focus\_level & 77.18 & $-$0.17 \\
\bottomrule
\end{tabular}
}
\caption{Quality ablation results. Each experiment clamps one feature to its minimum while optimizing the remaining features with NSGA-II.}
\label{tab:quality_ablation_appendix}
\end{table}

\section{Computational Cost Analysis}
\label{sec:cost_analysis_appendix}

We analyze the approximate per-query computational cost of FeatGEO with gpt-4o-mini as the backbone model ($P{=}8$, $G{=}8$ generations, $n{=}3$ fusion completions per evaluation). Because the implementation includes cache loading and reuse, these measurements are not exact and may vary with cache state and I/O overhead. As shown in Table~\ref{tab:cost_per_step}, the GA optimization stage still accounts for the majority of computation, representing 87.8\% of total wall time and 86.8\% of total prompt tokens.

\begin{table}[H]
\centering
\resizebox{\columnwidth}{!}{%
\begin{tabular}{lrrrr}
\toprule
\textbf{Pipeline Stage} & \textbf{Time (s)} & \textbf{API Calls} & \textbf{Prompt Tok.} & \textbf{Compl.\ Tok.} \\
\midrule
Feature Extraction & 17.8 & 5 & 19,011 & 740 \\
Initial Population & 192.2 & 41 & 113,318 & 16,495 \\
GA Optimization & 1,510.8 & 320 & 874,828 & 133,232 \\
\midrule
\textbf{Total (per query)} & \textbf{1,720.8} & \textbf{366} & \textbf{1,007,157} & \textbf{150,467} \\
\bottomrule
\end{tabular}
}%
\caption{Average per-query computational cost breakdown of FeatGEO (gpt-4o-mini).}
\label{tab:cost_per_step}
\end{table}

\end{document}